\documentclass{iaus}
\usepackage{graphicx}

\title[JD 11.~~Collision of Magnetic Flux Tubes] 
{Signature of Collision of Magnetic Flux Tubes in the Quiet Solar Photosphere}

\author[Aleksandra Andic]   
{Aleksandra Andic}

\affiliation{BBSO, 40386 N. Shore Ln.\\Big Bear City, CA-92314, USA\\ email: {\tt aandic@bbso.njit.edu} }

\pubyear{2010}
\volume{273}  
\pagerange{119--126}
\jname{Physics of Sun and Star Spots}
\editors{A.C. Editor, B.D. Editor \& C.E. Editor, eds.}
\begin{document}

\maketitle

\begin{abstract}
Collision of the magnetic flux tubes in the Quiet Sun was proposed as one of the possible sources for the heating of the solar atmosphere (Furusawa and Sakai, 2000). The solar photosphere was observed using the New Solar Telescope ad Big Bear Solar Observatory. In TiO spectral line at 705.68 nm we approached resolution of $0.1''$. The horizontal plasma wave was observed spreading from the larger bright point. Shorty after this wave an increase in the oscillatory power appeared at the same location as the observed bright point. This behavior matches some of the results from the simulation of the collision of the two flux tubes with a weak current. 
\end{abstract}

\firstsection 
\section{Introduction}

The heating of the solar atmosphere is not well understood. Through the years of the research various solutions for the heating of the atmosphere were suggested and tested. One of the suggestions for the possible energy supply is collision of the magnetic flux tubes in the Quiet Sun (Furusawa \& Sakai, 2000). Furusawa and Sakai model present these collisions as the source of the fast magneto acoustic waves. Those waves should form shocks higher up in the atmosphere and deposit the energy trough the dissipation of the shock front. \par 
The solar photosphere has ubiquitous magnetic field. This fact is confirmed with the observations (Orzoco Su\'{a}rez {\it et al.}, 2007; Lites {\it et al.}, 2008) and simulations (Sch\"{u}ssler and V\"{o}gler, 2008; Steiner {\it et al.} 2008). Because the magnetic field is ubiquitous the importance of its study increases in the the quiet Sun, as well. The New Solar Telescope (NST) at Big Bear Solar Observatory (BBSO) revealed previously unresolved small structures in the quiet Sun photosphere. Moreover, previously unresolved bright points turned out to be a source for the part of the oscillations that were previously exclusively contributed to the intergranular lanes (Andic {\it et al.} 2010).  We are now able to resolve those small structures and study their dynamic in detail. \par 

With NST at BBSO we were able to observe bright points in such detail that we're able to detect some of the signature proposed for collision of the magnetic flux tubes.

\section{Data and Analysis}

The data-set used in this analysis was obtained on 29 July 2009. We used NST at BBSO (Goode {\it et al.} 2010) and Nasymut optical setup that consists from TiO broadband filter and PCO.2000 camera. A filter has central wavelength at 705.68 nm with the bandpass of 1 nm (Andic {\it et al.} 2010). The target of the observations was the center of the quiet Sun disk. The data sequence consist of 120 speckle reconstructed images with the temporal cadence of 15 s. The performed image reconstruction is based on the speckle masking method of von der L\"{u}he (1993). To reconstruct images we used the procedure and code described in W\"{o}ger {\it et al.} (2008). \par 
The intensity oscillations were detected using the wavelet analysis (Torrence \& Compo, 1998) with the approach to the automated analysis described in detail in Andic {\it et al.} (2010). The speed was calculated using the nonlinear affine velocity estimator (NAVE) method which applies an affine velocity profile instead of a uniform velocity profile commonly used in the local correlation tracking (LCT) method (Chae \& Sakurai, 2008).

\section{Results}

The Doppler velocity maps, calculated from the intensity images of the quietest Sun region, showed that some of the BP have a ring-like speed distribution. Around 11\% of the total number of analyzed bright points had similar speed distribution. We speculate that this kind of distribution is connected with the collision of the magnetic flux tubes with the weak current (Furusawa \& Sakai, 2000). No circular ring-like distributions were observed; the speed distribution was distorted by the presence of the surrounding granules or nearby bright points. \par 
The analysis of the intensity oscillations of the BPs showed a spike in the oscillatory power appearing shortly after the ring-like speed distribution. This finding indicate the increased oscillatory emission. However, since we had only intensity images from one formation height in the photosphere, there was no reliable method to confirm what kind of oscillations we observed. Furusawa \& Sakai (2000) expected strong fast magnetosonic waves. \par
\begin{figure}[h]
\begin{center}
 \includegraphics[width=5.4in]{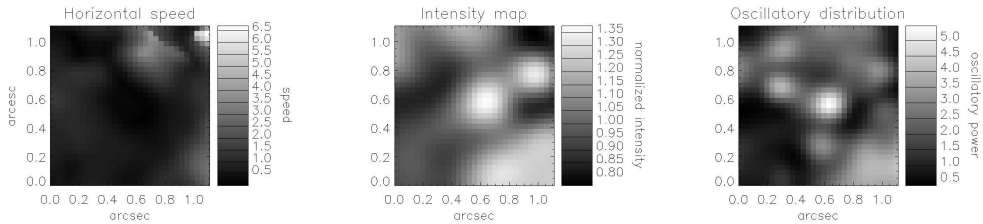} 
 \caption{The image shows an example of the signature of the collision of two parallel flux tubes with weak current. The panel marked 'Intensity map' shows the analyzed bright point. The two close bright points are visible with the analyzed one centered. Panel marked with 'Horizontal speed' shows the plasma flow we detected around these bright points. It is visible that there is ring-like increase in the speed of the plasma flow all around the analyzed larger bright point. Panel marked with 'Oscillatory distribution' shows the spatial distribution of the oscillatory power around and on top of these bright points. The largest power in the area is detected at the same location as the largest bright point.}
   \label{fig1}
\end{center}
\end{figure}

 
Fig.\,\ref{fig1} shows an example of the speed distribution. In this figure we have an example of two neighboring bright points from which only one shows this signature. The peak of the oscillatory power occurred few minutes after we observed the ring like distribution as visible at the Fig.\,\ref{fig3}.

\begin{figure}[h]
\begin{center}
 \includegraphics[width=5.4in]{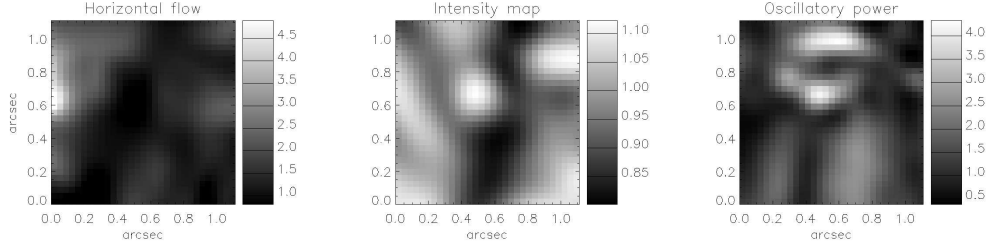} 
 \caption{Another example of the slightly deformed signature appearing around the single bright point. Panels illustrate the same quantities as in image above. }
   \label{fig2}
\end{center}
\end{figure}

\begin{figure}[h]
\begin{center}
 \includegraphics[width=3.4in]{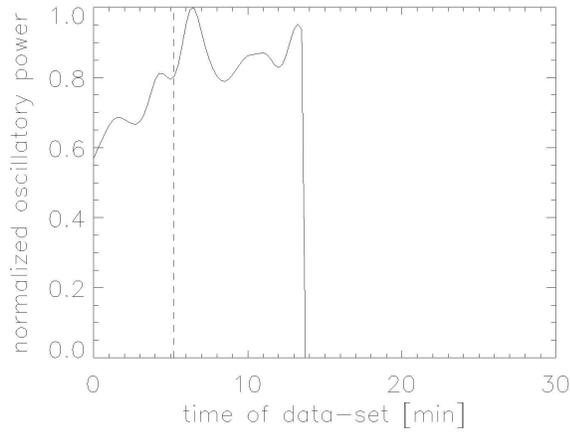} 
 \caption{The image shows a temporal change in integrated oscillatory power above the brightest bright point of the couple of the bright points shown at Fig.\,\ref{fig1}. Vertical dotted line marks a moment when the ring like signature was observed. It is noteworthy that after appearance of the ring-like speed distribution we have a significant increase in the oscillatory power emitted by the bright point. The graph illustrates only power registered during the lifetime of the BP.}
   \label{fig3}
\end{center}
\end{figure}

\begin{figure}[h]
\begin{center}
 \includegraphics[width=3.4in]{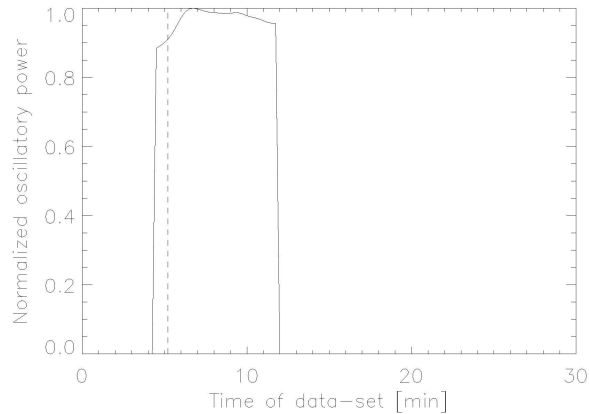} 
 \caption{The image shows a temporal change in integrated oscillatory power above the  bright point from the second example shown at Fig.\,\ref{fig2}.  Vertical dotted line marks a moment when the ring-like signature was observed.}
   \label{fig4}
\end{center}
\end{figure}

Fig.\,\ref{fig2} shows example of the horizontal plasma flow when the analyzed bright point is surrounded only by granules, while Fig.\,\ref{fig4} shows corresponding change in the maximum power for the same bright point. This lone bright point also demonstrate the plasma flow that has ring-like speed distribution. The deviation of the circular shape is caused by plasma flow influences of the nearby granules. 

\section{Conclusions}

The ring-like speed distribution around the bright point was followed closely in time by rise in oscillatory power detected at the same location at which was the bright point in question. This sequence of the events  might indicate that there was a collision of the magnetic flux tubes happening inside that bright point. Due to the limitation of our data set this conclusion is speculative. Our data set consist only from the intensity images, without any magnetic information nor information about line of sight velocities. Hence, this result needs to be tested with more complete data-sets.\par 
 Although we still cannot resolve the substructure of the bright point itself, this signature points to another speculation that inside of the individual BP  are individual flux tubes, indicating that the individual BP might be composed from the smaller, unresolved flux tubes.

\begin{discussion}

Discussion about this topic happened on one-to-one basis, and encompassed not only results showed here but also available and planed instrumentation at NST at BBSO. All questions and answers would cause this article to go over the page limit, hence discussion is omitted. The interested scientist were V.M. Pillet, D.S. Bloomfield, D. Tripathi and L. Bharti. 
\end{discussion}

\end{document}